\documentclass[letterpaper]{jpconf}
\usepackage{graphicx}
\begin{document}
\title{$\nu_{e}$ At T2K}

\author{Melissa George, on behalf of the T2K collaboration}

\address{Queen Mary, University of London}

\begin{abstract}

Tokai-to-Kamioka T2K is a long baseline neutrino oscillation experiment, looking for sub-dominant $\nu_{\mu} \rightarrow \nu_e$ oscillations.  One of the primary aims of the T2K experiment is to narrow down the current limit on the value of $\theta_{13}$ (which if this value large enough, suggests CP violation in the neutrino sector) and to find whether $\theta_{23}$  is maximal, which is crucial for constraining neutrino mass models.  T2K produces a high power neutrino beam at the J-PARC facility on the east coast of Japan, and this beam is then characterised by the near detector ND280 280 m from the start of the beam, the far detector (Super-Kamiokande), a 50 kton water Cherenkov detector, then detects the beam at the oscillation maximum of 295 km on Japan's west coast. T2K will be the first experiment to really study the $\nu_e$ appearance measurement---whose result will be sensitive to $\theta_{13}$ arguably the main physics goal of T2K.  The ND280 detector is imperative to this measurement and will be used to understand the $\nu_e$ appearance background.  The status of the T2K experiment and the predicted performance for the $\nu_e$ appearance measurement is presented here.
\vspace{-12mm}
\end{abstract}

\section{Introduction}

The study of neutrino oscillations is one of the most active and promising experimental fields of particle physics research.  They are the first direct evidence of physics beyond the standard model, which has, until now, proved reliable and a good approximation at low energies, and have paved the way for new physics and one day, for a new model of particle physics that holds at higher energies.  Neutrinos have three mass states $\nu_{1,2,3}$, and their mixing is described by three independent mixing angles $\theta_{12,23,13}$, two independent signed mass-squared differences $\Delta m^{2}_{12,23}$ and one CP-violating phase angle $\delta$.  The two important parameters that will be measured are the disappearance of $\nu_\mu$ and the appearance of $\nu_e$.  We can simplify their probabilities since we can define the effective mixing angles and $\nu_\mu$ disappearance experiments have led to constraints, namely $\sin^2 2\theta_{\mu} > 0.89$ and $1.6 \times 10^{-3} < \Delta m^2_{23} < 4 \times 10^{-3} \mathrm{eV}^2$.  For $\Delta m^2_{23} \simeq 6 \times 10^{-3} \mathrm{eV}^2$, $\sin^2 2\theta_{13} < 0.05$ and for $\Delta m^2_{23} \simeq 2 \times 10^{-3} \mathrm{eV}^2$, $\sin^2 2\theta_{13} < 0.12$ at $90\%$ C.L\cite{Fukuda:1998mi}.  Since atmospheric neutrino data indicates almost full mixing $\theta_{23} \simeq \frac{\pi}{4}$ and since $\theta_{13}$ is so small, the probabilities are:

\begin{equation}
  P ( \nu_\mu \rightarrow \nu_\mu ) = 1 - \left(\sin^2 2\theta_{23}  \sin^2 \left(\frac{1.27 \Delta m^2_{23} ( \mathrm{eV}^2 )}{\frac {E ( \mathrm{GeV}/c^2 )}{L ( \mathrm{km} )} }\right)\right)
\end{equation}

\begin{equation}
  P ( \nu_\mu \rightarrow \nu_e ) = \frac{1}{2}\sin^2 2\theta_{13} \sin^2 \left(\frac{1.27 \Delta m^2_{23} ( \mathrm{eV}^2 )}{\frac {E ( \mathrm{GeV}/c^2 )}{L ( \mathrm{km} )} } \right). 	
\end{equation}

Hence we can see that the disappearance of $\nu_\mu$ leads to the determination of $\theta_{23}$ and $\Delta m^2_{23}$ while the appearance measurement can be used to measure $\theta_{13}$.  Measurements of $\nu_\mu$ disappearance have already been made using atmospheric neutrinos at Super-K\cite{Fukuda:1998mi} and accelerator neutrinos, but will be improved by T2K, while the appearance of $\nu_e$ would be observed for the first time.  For the full treatment we refer to earlier in these proceedings\cite{LLWI2010:Dean}.  We have yet to measure directly the value of $\theta_{13}$, nor to understand the mass hierarchy of neutrinos.  T2K aims to answer these questions.
\vspace{-1mm}

\section{$\theta_{13}$: $\nu_{\mu}\rightarrow\nu_{e}$ Signal And Background}

In order to observe and measure $\nu_{\mu}\rightarrow\nu_{e}$ oscillations, one must be able to believe the signal and understand the backgrounds as fully as possible.  The primary experimental concerns of such a measurement are the high level of background and the statistical limitations on all but the largest currently allowed values of $\theta_{13}$ (this can be minimised by maximising the neutrino flux---see the following sections).  

\begin{figure}[!h]
 \begin{center}
  \begin{tabular}{cc}
   \begin{minipage}{0.4\textwidth}
	\includegraphics[width=1.5in,height=1in,keepaspectratio=true,trim = -20mm 0mm 0mm 0mm]{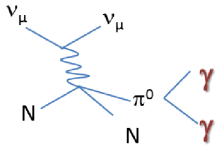}
	\caption{\footnotesize The NC$\pi^0$ background process to the $\nu_e$ appearance measurement.}
	\label{fig:nuePiBkgrdFeynman}
   \end{minipage}
   &    
   \begin{minipage}{0.558\textwidth}
The background events to $\nu_e$ appearance can be split fundamentally into two categories, each accounting for 50\% of the total background.  These are: the intrinsic $\nu_e$ contamination of the beam and NC-1$ \pi^0$ events that can be misidentified as electrons from electron neutrinos.  For $\nu_e$ appearance measurements a common example of this $\pi^0$ background can be seen in figure \ref{fig:nuePiBkgrdFeynman}.  For further information regarding this $\pi^0$ background we refer to elsewhere in these proceedings\cite{LLWI2010:Ben}.
   \end{minipage}
   \end{tabular}
 \end{center}
\end{figure}
\vspace{-6mm}

\section{T2K's Sensitivity To $\nu_e$ Appearance and Hence to $\theta_{13}$}

The current best limit on the value of $\theta_{13}$ was made by the CHOOZ reactor experiment nearly ten years ago\cite{Apollonio:1999ae}.  The T2K experiment, with over 500 collaborators from 62 institutions in 12 countries, can improve this limit by an order of magnitude, using the worlds highest power neutrino beam, an off-axis beam geometry and a set of detectors designed to well characterise the $\nu_e$ appearance background.

\begin{figure}[!h]
 \begin{center}
  \begin{tabular}{cc}
   \begin{minipage}{0.68\textwidth}       
    \includegraphics[height=1.8in,keepaspectratio=true,trim = 0mm 0mm 0mm 0mm]{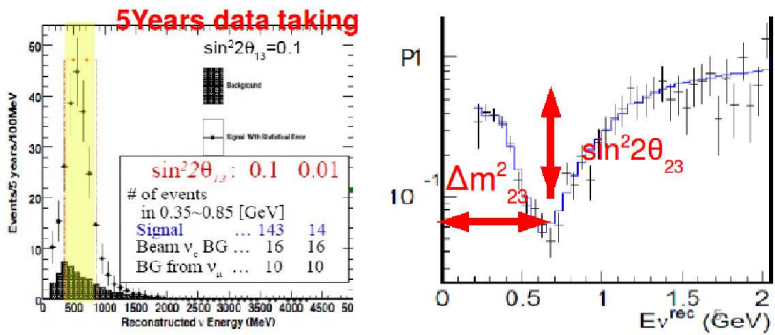}
\label{fig:bckgrdVsSignal}
   \end{minipage}
   &    
   \begin{minipage}{0.29\textwidth}
   \caption{\footnotesize Signal vs. background for the $\nu_e$ appearance measurement at T2K over $8 \times 10^{21}$ POT at $30 \mathrm{~GeV}/c^2$ at the oscillation maximum and the dependence of $\sin^{2}2\theta_{23}$ on the probability of oscillation, showing the importance of observing the oscillation at its maximum and the high signal to background ratio for the $\nu_e$ appearance measurement.}
   \end{minipage}
   \end{tabular}
 \end{center}
\end{figure}

From figure \ref{fig:bckgrdVsSignal} we can see that since the value of $\sin^{2}2\theta_{23}$ depends on the probability of oscillation, it is essential to focus our detection capabilities at the oscillation maximum and also that at this oscillation maximum the expected signal at T2K is far greater than the background over five years data taking or $8 \times 10^{21}$ POT at 30~GeV, thereby making this one of the primary experimental concerns of the T2K experiment.  This is achieved through careful consideration of the T2K beam.  The positioning of the near and far detectors at an off-axis beam geometry and the distance between the near and far detectors at the oscillation maximum of 295 km.

\section{The T2K Neutrino Beam}

The T2K neutrino beam is essential to T2K's ability to measure $\nu_{\mu}\rightarrow\nu_{e}$ oscillations, since $\nu_e$ contamination in the beam accounts for 50\% of the $\nu_e$ appearance background.  The proton synchrotron at J-PARC uses dual purpose (dipole and quadrupole) super conducting magnets to bend the proton beam into a small radius, (necessary due to the lack of space at J-PARC).  The beam is then separated from the target station and collides with a helium cooled graphite target to produce positively charged pions of equivalent energy to the incoming protons.  These pions are then focused by a horn and enter a 100~m helium filled decay volume where a large proportion will decay into muon neutrinos.  Any remaining protons and pions will be captured by the beam dump.  The direction of the beam will be determined on a spill by spill basis by muon sensors that are positioned behind the beam dump.  The beam created is the highest power pulsed neutrino beam in existence, it will produce 0.75MW of protons on target leading to 4 MW after five years.  This will create the maximum possible neutrino flux and increase the statistical sensitivity of T2K, which is  imperative to an experiment looking for sub-dominant oscillations.  

\begin{figure}[!h]
 \begin{center}
  \begin{tabular}{cc}
   \begin{minipage}{0.6\textwidth}
	\includegraphics[height=2.3in,keepaspectratio=true,trim = -20mm 0mm 0mm 0mm]{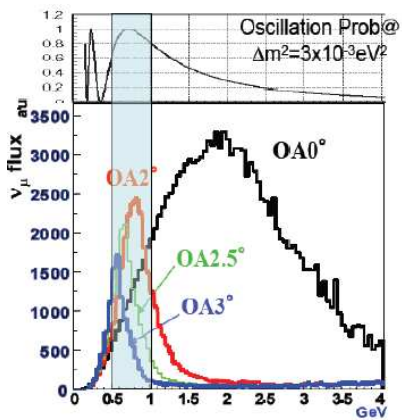}
	\caption{\footnotesize The neutrino flux against the neutrino energy for beam angles between $0^{\circ}$ and $3^{\circ}$.  The 1st oscillation maximum is highlighted to show the peak of the spectra.  We can see that a beam at $2.5^{\circ}$ is the most highly focused around the oscillation maximum, the flux is highly peaked and that there is very little high energy tail.}
	\label{fig:beamGeom}
   \end{minipage}
   &    
   \begin{minipage}{0.38\textwidth}
It is necessary to maximize the $\nu$ flux at energies corresponding to the oscillation maxima whilst minimising the high energy tail and the electron neutrino contamination in the beam.  To do this an off-axis beam geometry of $2.5^{\circ}$ is used, providing a (quasi) mono-energetic beam tuned to the first oscillation maximum, increasing the flux and drastically cutting the high energy tail responsible for inelastic interactions that are background to the CCQE events (see figure \ref{fig:beamGeom}).  The $\nu_e$ contamination of the beam is also minimised around the peak producing a beam composition of around 95\% $\nu_\mu$, 4\% $\overline{\nu}_\mu$ and less than 1\% $\nu_e$.
   \end{minipage}
   \end{tabular}
 \end{center}
\end{figure}
\vspace{-5mm}

\section{ND280 and the $\nu_e$ Appearance Measurement}

There are two near detectors situated at 280 m from the start of the beam: the INGRID on-axis detector that measures the exact position of the neutrino beam (taking measurements at least daily, often hourly) and the ND280 off-axis detector.

The ND280 off-axis detector consists of a suite of eight near detectors, in the same pit as the INGRID, but in the off axis position---in line with Super-Kamiokande.  The eight detectors are the Pi0 detector (P0D), the Time Projection Chambers (TPCs) and Fine Grain Detectors (FGDs), which collectively are known as the Tracker, the Barrel-, P0D- and Downstream-Electromagnetic Calorimeters (ECals) and the Side Muon Range Detector (SMRD).  These detectors are then surrounded by a magnet producing a 0.2~T magnetic field.

$3.34\times 10^{3}$ $\nu_e$ events per $30\mathrm{~GeV}/c^2$ $1.6\times10^{21}$ POT (around one years data taking) are expected at ND280.  Approximately 4\% of these events produce $\pi^{0}$s (see figure~\ref{fig:nuePiBkgrdFeynman}) and around 20\% of these $\pi^{0}$s produce two $\gamma$s in the barrel and downstream ECals.

The $dE/dx$ measurement in the TPCs, which has been demonstrated in a test beam, makes the $\nu_e$ beam contamination measurement possible\cite{LLWI2010:Kendall}, whilst the ECal will measure the $\gamma$ pairs from single $\pi^{0}$'s produced by neutrino interactions in the inner detectors, as well as identifying $\mu$'s and $e$'s to help determine the $\nu_e$ contamination of the beam. 
\vspace{-2mm}

\section{$\nu_e$ Appearance Events at Super-K}

There are 143 $\nu_e$ appearance events expected at T2K for $0.75\mathrm{~MW} \times 5\times10^{7}$s, the expected running time of T2K.  Ten of these events will be background from $\nu_\mu$ events and 16 from beam $\nu_e$ events.  $\nu_e$ events are selected to have a vertex $> 2$~m from the tank wall in a direction of $< 25^\circ$ of the beam, with less than 16 hits in the outer tank, displaying a single $e$-like ring with no decay $e$ and $0.35 \mathrm{~GeV}/c^2 < E < 0.85 \mathrm{~GeV}/c^2$.  For more information see\cite{LLWI2010:Ben}.
\vspace{-2mm}

\section{The Experimental Readiness of T2K}

\begin{figure}[!h]
 \begin{center}
  \begin{tabular}{cc}
   \begin{minipage}{0.5\textwidth}       
    \includegraphics[height=2in,width=3in,keepaspectratio=true,trim = 0mm 0mm 0mm 0mm]{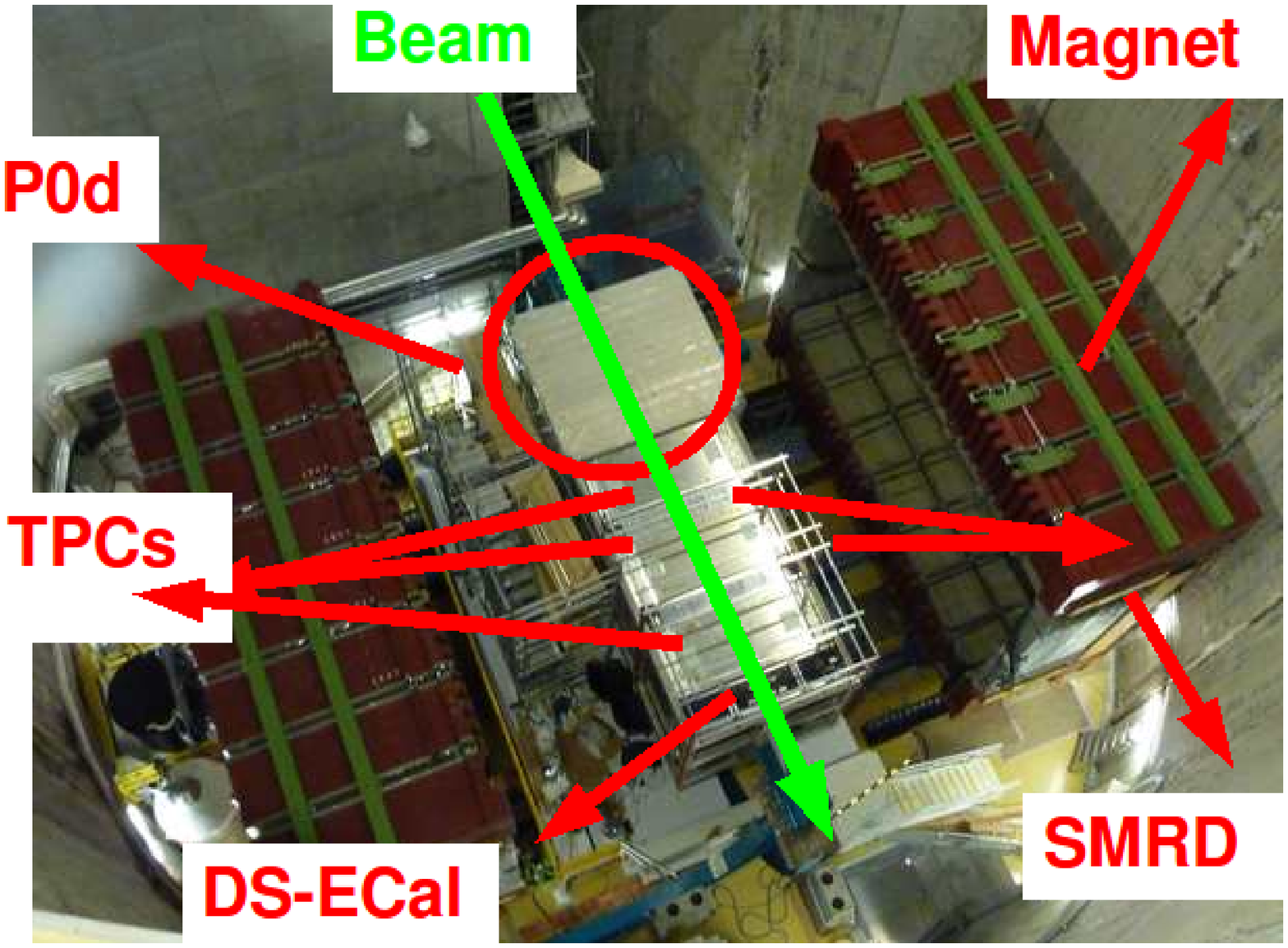}
    \caption{\footnotesize The ND280 Pit with all subdetectors (bar the barrel and P0D ECal modules that will be installed in the summer) installed, as of January 2010.}
	\label{fig:Pit}
   \end{minipage}
   &    
   \begin{minipage}{0.5\textwidth}
   \includegraphics[width=3in,height=2in,keepaspectratio=true,trim = 0mm 0mm 0mm 0mm]{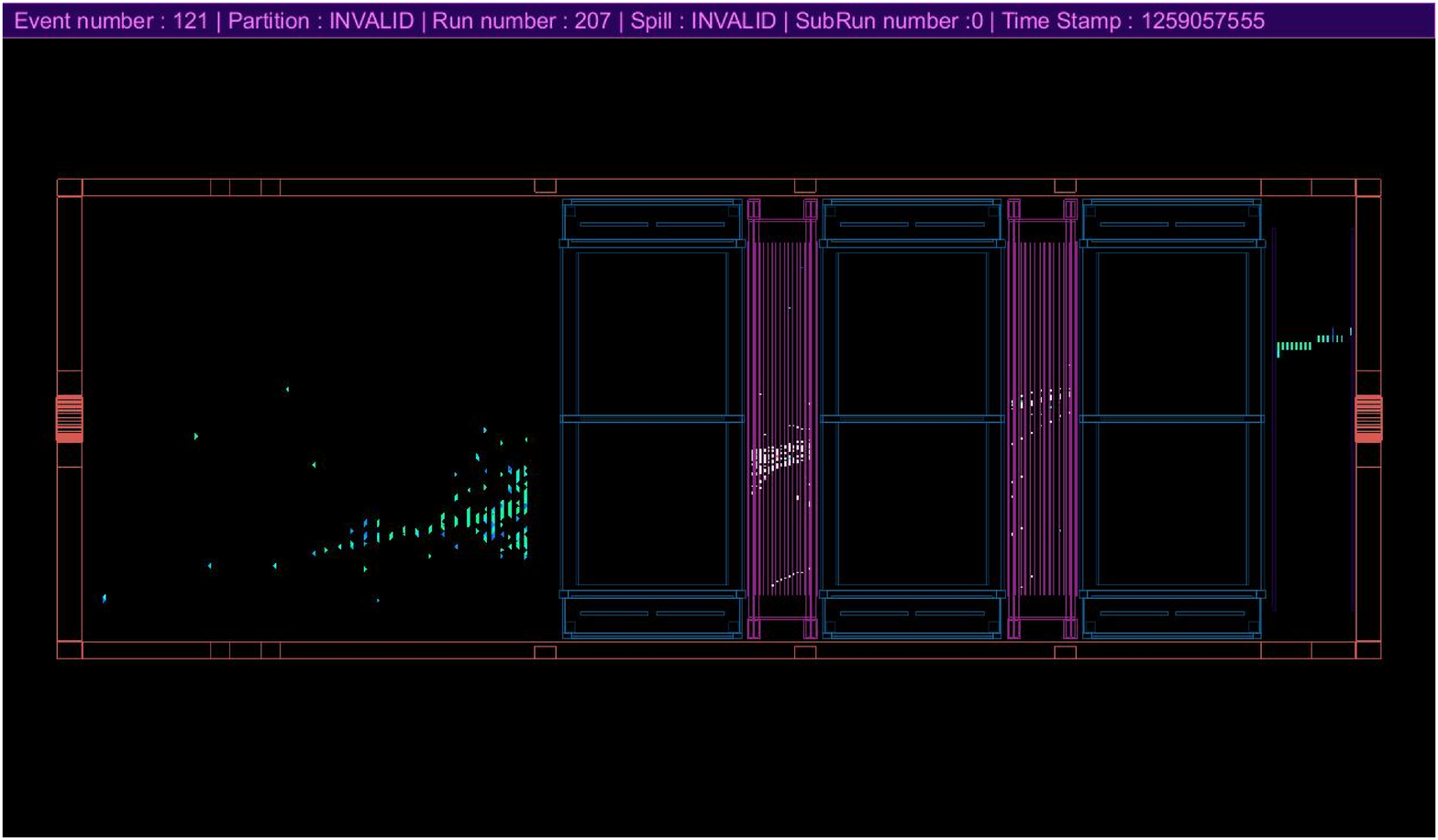}
   \caption{\footnotesize A $\nu_\mu$ event observed by the ND280 detectors.}
   \end{minipage}
   \end{tabular}
 \end{center}
\end{figure}

All ND280 subdetectors are fully commissioned, tested and operational, with the exception of the barrel and P0D ECals that are under construction and will be installed in summer 2010 as planned.  T2K has been taking data since late 2009 and is currently taking physics data. 
\vspace{-2mm}

\section{Conclusions}

T2K is the first neutrino oscillation experiment with discovery potential for $\nu_{\mu}\rightarrow\nu_{e}$ oscillations, expecting to observe over 100 such events (after background subtraction) over the course of its data taking.  The near on and off axis detectors, and the far detector Super-Kamiokande, are fully commissioned operational and data taking and all are already recording neutrino events.

\section*{Bibliography}
\bibliographystyle{iopart-num}
\bibliography{LLWIproceedings}

\end{document}